# Easily Adaptable Complexity Measure for Finite Time Series

Da-Guan Ke (柯大观)[1, 2, *] and Qin-Ye Tong (童勤业)[2]

[1]Department of Mathematics, Zhejiang University, Hangzhou 310027, PRChina
[2]Department of Biomedical Engineering, Zhejiang University, Hangzhou 310027, PRChina

We present a complexity measure for any finite time series. This measure has invariance under any monotonic transformation of the time series, has a degree of robustness against noise, and has the adaptability of satisfying almost all the widely accepted but conflicting criteria for complexity measurements. Surprisingly, the measure is developed from Kolmogorov complexity, which is traditionally believed to represent only randomness and to satisfy one criterion to the exclusion of the others. For familiar iterative systems, our treatment may imply a heuristic approach to transforming symbolic dynamics into permutation dynamics and vice versa.

## I. INTRODUCTION

Over the past 30 years, various complexity measures have been proposed and applied in many fields including neurology, physiology, and biology [1]. Most working measures follow three competing criteria. Such criteria, denoted as I, II, and III, are sketched by a monotonically increasing function, a convex function, and a monotonically decreasing function of "disorder" (randomness), respectively [2]. Although criterion II (the "one-hump" criterion) seems more intuitively accepted [3], it is still a long way to reach a consensus. From a practical viewpoint, a surprising and interesting compromise is to provide a single measure adaptable to two or more criteria. One such measure is $\Gamma_{\alpha\beta}$, defined in Ref. [2].

On the other hand, considerable progress has been made recently in partitioning real-world time series. *Permutation entropy* (PE) [4] was presented by Bandt and Pompe. The main advantage of PE over other entropies [5] is the natural quality of the permutation method, by which we should denote neighboring points according to their relative time order and then permute the denotations to the points' numerical order. However, the equal values are unmanageable. Such values in either periodic or non-periodic cases had simply been neglected. One who uses small random perturbations to break the equality cannot limit the introduced distortion to a tolerable level when the number of equal values increases.

Here we propose an adaptable complexity measure, *lattice complexity* (denoted by $C_L$),

---
[*] Email address: kdg@zju.edu.cn

based on Kolmogorov complexity [6], or more specifically on the algorithm of *Lempel-Ziv complexity* (denoted by $C_{LZ}$) [7], which has been effectively normalized [8] and widely applied [9]. We also propose a modified permutation method that naturally processes the equal values and significantly facilitates the transformation between the system-dependent symbolic sequence and the data-dependent *permutation sequence* (or *permutations* for short). This transformation implies an applied *permutation dynamics* other than the well-known symbolic dynamics to describe dynamical systems.

At one extreme $C_L$ is approximately equivalent to $C_{LZ}$ serving as a measure of criterion I [10]; while at the other extreme $C_L$ can qualify any sequence as the simplest. In a range between these two extremes, we will see that our measure meets criterion II. Unlike $\Gamma_{\alpha\beta}$ and its variants [11], our method enables us to count both the periodic matters and the completely chaotic matters as exactly the simplest objects. The last sequence taking the minimum $C_L$ is the sequence corresponding to the period-doubling accumulation points or the so-called "edge of chaos" [12] that are thought to lead to the most complex behavior [13]. It is very strange for a Kolmogorov-complexity-based measure to have such properties since Kolmogorov complexity is regarded as a definition of the randomness, the quality of possessing un-predictable disorder representing only criterion I [14].

Subtracting a criterion-I measure from a selected constant may produce a particular criterion-III measure and vice versa, as the "order" and "disorder" defined in Ref. [2]. In the following, we choose to pay more attention to the first two criteria instead of the third.

## II. LATTICE COMPLEXITY

According to Kolmogorov and others [5,11], the length of the shortest program that produces a given sequence should be viewed as its complexity measure. Kolmogorov complexity is not computable, because the program depends on the (Turing) machine you use and it is still impossible to ensure that the shortest program can be obtained. As a rough upper bound of Kolmogorov complexity, $C_{LZ}$ [7] is an easily calculable algorithm allowing only two operations: replication and insertion. This algorithm scans the sequence from left to right for a *new pattern*—i.e., the sub-sequence that cannot be generated by replication alone, but can only be completed by an insertion as the last letter. In other words, the new pattern except its last letter has a prototype in the *exhaustive history* prior to the pattern's penultimate letter, but the new pattern as a whole has no prototype in the exhaustive history prior to the pattern's last letter. The number of all new patterns is counted as the outcome of $C_{LZ}$ (see the example below). $C_{LZ}$ acts like entropy or a *Lyapunov Exponent* [15] so that meets criterion I.

Our algorithm is a little more complicated but less time-consuming than $C_{LZ}$ because of an additional operation offering further computational benefits. This operation should be thought as the "iterative map" on a certain set of words. With one word being mapped onto another repeatedly, a sequence composed of entirely distinct (or recurrent) words will be regarded as a chaotic (or periodic) iteration of the map [10].

Let us give a brief explanation. It is well known that symbolic dynamics provides a natural way to represent the complicated behavior of chaotic systems, especially when the coarse-graining procedure is based on the generating partition that defines distinct symbols. For a uni-modal interval map, only one critical point divides the whole interval into two segments labeled with "0" and "1." When the map is surjective, there is no forbidden sequence. The symbolic dynamics of the map is semi-conjugate to the Bernoulli-2 shift and every semi-infinite sequence is directly related to a unique orbit. Thus, every finite word corresponds to a particular segment covering a particular point visited by the orbit: the longer the length of the word, the smaller the width of the segment (see page 216 and 217 of Ref. [16] or Appendix A).

Now we assume that every word of a given length $r$ is allowed to be viewed as a point in an orbit. Since all orbits are sorted into the two classes—the chaotic and periodic—we will see that all sequences of words corresponding to distinct orbits are also sorted into the chaotic and periodic. Repeated words can only be found in periodic sequences, and the words within exactly one period or within a chaotic sequence must be distinct from each other. Note that all orbits of a simple map (e.g. logistic map) are produced with the same program of short length. We call a periodic or chaotic sequence an *iterative sequence* and deem any iterative sequence an object of plainness, not of complexity.

Combining the two operations of $C_{LZ}$ with the iteration, for any given sequence, we define a *lattice* as a sub-sequence with the following properties.
(a) The lattice begins with an iterative sequence that extends with the iteration until a symbol makes the sequence neither chaotic nor periodic.
(b) After the iteration has been completed, the lattice extends with the replication until the insertion is required.

Then we reckon the amount of the lattices as the complexity measure $C_L$ of the sequence.

Let us take an example. Suppose there is a binary sequence
$$s = 10001100101011.$$
Assuming that the word length $r$ equals 1, $s$ can be parsed following the two approaches $C_{LZ}$ and $C_L$ represented by the dot and the mark "$\vee$," respectively:
$$s = 1 \cdot 0 \cdot 001 \cdot 1001 \cdot 01011 \cdot ,$$
$$s = 10001 \vee 1001 \vee 01011 \vee .$$

Following the procedure of $C_{LZ}$, the first two digits 10 are inserted each supplemented by a dot, because any of the digits has not appeared before. With a pre-existence in its exhaustive history 10, the third digit 0 is produced by copy. Without being interrupted by a dot, the fourth digit must be considered with the third together. Since these two digits 00 can be replicated from the exhaustive history 100 prior to the fourth, there is still no dot added immediately. However, the fifth digit 1 should be inserted and be accompanied by a dot, for the pattern 001 has no template in the history 1000. Having inserted digits, the sub-sequences 10001 and 1001 are both new patterns. The last five digits that cannot be produced by replication compose a complete new pattern.

By using our algorithm, any iterative sequence should always be assumed as a chaotic sequence at first. The first two digits 10 do not negate the assumption, but the third and the fourth 00 show that the sub-sequence 1000 is a fixed-point sequence (of period 1) with an initial state 1. The fifth digit fails to be fixed to 0, indicating the first iterative sequence has merely four digits. Because there is no predecessor that can be copied, the first five digits 10001 form the foremost lattice as a complete one. The mark "$\vee$" means that the previous digit is inserted and that a new lattice has started from the null sequence $\Lambda$. The following subsequence 1001 is another lattice that cannot be reproduced from the exhaustive history 10001100. The sub-sequence 0101 is a periodic sequence of period 2, but the iteration is interrupted by a following digit 1 which does not match the periodic regularity. Since there is no original prototype for 01011 in the history, this sub-sequence is a complete lattice suspended by a mark "$\vee$." We count the number of lattices as the number of insertion marks plus 1.

In this case, we see that $C_{LZ}(s) = 5$ and $C_L(s) = 4$.

When we set the word length $r = 2$, simply with a normal binary-quaternary table we have a *refined alphabet* $S^2 = \{0,1,2,3\}$. The *refined sequence* $s^2 = 2001320121213$ translated from the original sequence $s$ should be parsed as

$$s^2 = 2 \cdot 0 \cdot 01 \cdot 3 \cdot 201 \cdot 21 \cdot 213 \cdot,$$

$$s^2 = 2001 \vee 320121 \vee 213.$$

Then we see that $C_{LZ}(s^2) = 7$ and $C_L(s^2) = 3$. More results such as $C_{LZ}(s^3) = 9$, $C_L(s^3) = 2$, $C_{LZ}(s^4) = 10$, $C_L(s^4) = 2$, $C_{LZ}(s^5) = 9$, and $C_L(s^5) = 1$ are obtained similarly.

We call the increasing process of the word length the *fine-graining process* and the word length *r* the *fine-graining order*. In the above instance, when $r \geq 5$, $C_L$ remains 1. We say that the *critical order* $r^*$ of $s$ is 5, or $r^*(s) = 5$.

The difference between $C_{LZ}$ and $C_L$ enlarges with the fine-graining order unless the critical order has been reached. The reason lies in the fact that the fine-graining process leads to a drastic increase both in number and in length of the possible iterative sequences. It is clear that a chaotic sequence, or a one-period subsequence of a periodic sequence, cannot have a length longer than the size of the alphabet. Considering the sequences of the length equal to the size of the alphabet, there are at most $\alpha!$ different chaotic sequences of length $\alpha$ for a basic alphabet $S$ with size $\alpha$, $(\alpha^2)!$ sequences of length $\alpha^2$ for a refined alphabet $S^2$ with size $\alpha^2$, $(\alpha^3)!$ sequences of length $\alpha^3$ for $S^3$ with size $\alpha^3$, and so on. More details of the two measures' behavior concerning the fine-graining process have been discussed in Ref. [10]. Here we recall the following essential propositions.

Given the original sequence $s$ of length $n$, let $C_{LZ}(r,n)$ and $C_L(r,n)$ denote $C_{LZ}$ and $C_L$ of order $r$, respectively.

**Proposition 1.** When $s$ is an *m*-periodic sequence ($m < n$) and $r \geq m$, it is true that

$$C_{LZ}(r,n) = m,$$

$$C_L(r,n) = 1.$$

Here the *m*-periodic sequence is not the above-mentioned periodic sequence that belongs to the iterative sequence. In the *m*-periodic sequence, not all symbols within one period need to be different from one another. However, the refined sequence with an order $r \geq m$ is a periodic sequence.

**Proposition 2.** Suppose $S$ is the basic alphabet of $\alpha$ letters. If each letter in $S$ has the uniform probability $\alpha^{-1}$ to appear, for $r \sim n$ we obtain

$$\lim_{n \to \infty} \lim_{r \to \infty} \Pr[C_{LZ}(r,n) = n - r + 1] = 1,$$

$$\lim_{n \to \infty} \lim_{r \to \infty} \Pr[C_L(r,n) = 1] = 1.$$

As a **corollary**, when *n* is a constant number, we have

$$\lim_{r \to n} \Pr[C_{LZ}(r) = n - r + 1] = 1,$$

$$\lim_{r \to n} \Pr[C_L(r) = 1] = 1.$$

We see that there exists a critical order $r^* < n$ of $s$ with probability 1. When fine-graining order is $r \geq r^*$, $C_L$ remains 1 and qualifies $s$ as the simplest sequence, while $C_{LZ}$ remains the upper bound $(n - r + 1)$ and qualifies $s$ as the most complex sequence.

For a given alphabet with size $\alpha$, a *De Bruijn sequence* is an $\alpha$-ary sequence that every possible word of length $k$ occurs just once.

**Proposition 3.** If $s$ is a *De Bruijn sequence* of length $n = \alpha^k + k - 1$ ($k \in N$), we see

that

$$r^*(s) = k.$$

For example, the sequence 00110 is a binary De Bruijn sequence of order $k = 2$, in which every possible 2-bit word appears once. De Bruijn sequences are counted as approximations of completely random sequences [7]. Among the non-periodic sequences that have equiprobable letter distributions, De Bruijn sequences of length $n$ take the smallest critical order less than $\log_\alpha n$, meaning that a relatively low order (about $\log_\alpha n$) is enough to guarantee that De Bruijn sequences are accepted as the simplest by using $C_L$.

Roughly speaking, the higher randomness a non-periodic sequence shows, the lower critical order it has. Concerning every specific chaotic system, the symbolic sequence of a completely chaotic orbit will have a comparatively low critical order so that such a sequence may easily be viewed as the simplest case, as well as that of the periodic orbit. Only sequences that have both regularity and randomness have grave difficulties to be qualified as the simplest with the minimal outcome of $C_L$ as 1. An extreme instance with a much larger critical order is a sequence related to the period-doubling accumulation point. The critical order can even be half of the sequence length because every bifurcation makes the period of the corresponding symbolic sequence double once (see pp. 122 and 123 of Ref. [18]). When an appropriately large order has been used, a period-doubling sequence will take the outcome of $C_L$ higher than others take. However, since $C_L$ is relative and adaptable, there is no absolute boundary between chaotic sequences and those of "edge of chaos."

An iterative sequence is a sequence that can be generated by a single successive iteration of a deterministic map. Since a surjective logistic map has no forbidden sequence (see Appendix A, Ref. [16] or [18]), any sequence can be generated by a successive iteration of this map. Of course, for a given sequence, many other maps can also produce it. When the map has been determined, what makes the difference is how precise the description of every point is needed in the calculation to create the sequence. The precision of every point is represented by the fine-graining order. Finding both a proper precision and a proper map to ensure the program is shortest seems still impossible. Fortunately, when the fine-graining order has been used as a control parameter, for a rough estimation algorithm of Kolmogorov complexity, it is enough to find how many non-overlapping sub-sequences each can be generated by a single iteration. We do not need to detect any exact map corresponding to an iterative sequence, because any iterative sequence has already been assumed to be plain despite what map should be used.

Because of the existence of the iterative sequence, $C_L$ does not need to retrospect the exhaustive history of every letter so that $C_L$ can be calculated faster than $C_{LZ}$.

# III. PERMUTATION METHOD

For the real-world time series, the permutation method of PE [4] is a plausible choice. When the generating dynamics is unknown and locating the generating partition is impossible, it is still workable to compare all the values with their neighboring points to obtain the permutations rather than with the dynamical critical points to obtain the symbolic sequence. Due to our modifications, the equal values that cannot be managed by the original method now can be dealt with naturally.

Suppose we have a series of eight values:

$$x = \{2,4,8,4,3,6,6,7\}.$$

When the embedding dimension $d$ is 2, seven pairs of neighbors are organized. When $d$ is 3, six triples are arranged. For any $d < 8$, there are $8 - d + 1$ words of $d$ digits. Following the method of PE, since we have to add small random perturbations on the equal values within a word of length $d$, we will obtain different permutations by probability every time [4]. This will not happen in our measure. To get any $d$ permutation, we create an alphabet $\{0,1,...,d-1\}$ and denote every point by an element of this alphabet. More precisely, we denote the lowest value by 0, the second to the lowest by 1, and so forth. If there are $m$ equal values, we denote them exactly by the lowest one among the $m$ possible symbols and remove the other $m-1$ symbols from the alphabet. Then the time series $x$ is transformed into the permutations as

$$P(d = 2, x) = (01, 01, 10, 10, 01, 00, 01),$$
$$P(d = 3, x) = (012, 020, 210, 102, 011, 002).$$

We obtain *permutation* $C_L$ of $x$ as

$$C_L^p(2, x) = C_L(P(2,x)) = 2,$$
$$C_L^p(3, x) = C_L(P(3,x)) = 1.$$

When the time series $x$ takes the place of the symbolic sequence $s$, the embedding dimension $d$ substitutes the order $r$, and $d!$ replaces $\alpha$, there are two other propositions similar to Propositions 1 and 2 being valid. For simplicity, we call both $d$ and $r$ *refining order*. There exists another critical order $d^*$ such that if $d \geq d^*$ then $C_L^p(d, x) = 1$.

Proposition 3 is applicable only in the rare case that the time series $x$ itself or its permutations is a De Bruijn sequence. However, Proposition 3 can also be converted into a new proposition more fitted to permutations when the De Bruijn sequence is displaced by another kind of artificial sequence. With a parameter $l$ doing the similar job of $k$, such a sequence has the length of $n = l!+l-1$ and all the numbers are constructed in a way that every $l$-permutation appears once. Thus, all $l!$ permutations are contained in the sequence and its critical order is $l$. Although the theoretical properties of this kind of sequence deserve to be well investigated, in this paper we shall confine our attention to the behavior of complexity measures. Note that this sequence and De Bruijn sequence characterize two diverse but similar qualities of the randomness: one has equiprobable permutations and the other has equiprobable sub-sequences. We call both kinds of sequences *emblematic sequences* of randomness.

Extensive information of these new propositions is given in Appendix B.

For a wide range of interval maps, with our modifications one can take permutations from not only chaotic orbits, but also periodic orbits, and see that using a modified permutation method [17] is still "similar to using generating partitions" [4]. Furthermore, when permutations can be ordered by some simple conventions as those for symbolic sequences (see Appendix A), the transformation between the permutations and the symbolic sequences will be carried out freely despite the possible existence of equal symbols or words.

Considering the shift (or doubling) map $x_{t+1} = 2x_t \pmod 1$, we divide the interval into the two segments (0,0.5) and (0.5,1) denoted by 0 and 1, respectively. Symbolic sequences can be ordered as if they are binary integers [18]. With a given order $r$, the sequences can be translated into integer sequences with a precision of $2^r$ and then can be transformed into permutations of order $d$. Most $d$-permutation sequences of an orbit can also be transformed into symbolic sequences. Because $x_t < x_{t+1}$ for $0 < x_t < 0.5$ and $x_t > x_{t+1}$ for $0.5 \le x_t < 1$, by comparing every pair of permutation symbols within a $d$ permutation, one can judge what monotonic segment the former symbol connects and transform every permutation symbol except the last one into to a symbolic symbol. For instance, a permutation sequence (132,321,312) of order 3 can be transformed into a symbolic sequence as 0110.

The relationship between permutations and symbolic sequences may imply a practical permutation dynamics. This issue will be discussed in a separate paper. For a more theoretical permutation dynamics, we refer to a broader theory called "combinatorial dynamics" [19].

## IV. EXAMPLES

As an example, we created a pseudo-random time series $x$ by simple multiplicative congruential method [20] and then the binary symbolic sequence $s$ by the median partition of $x$ (for $x_i \in [0,1]$, $s_i = 0$ if $x_i < 0.5$ and $s_i = 1$ otherwise). In order to compare the results with that of De Bruijn sequence conveniently, we took the length $n = 2^k + k - 1 = 8204$ ($k = 13$). It was found that $r^* = 25 > k$ and $d^* = 12 < k$ (Fig.1). Actually, the critical orders $r^*$ and $d^*$ varied slightly in repeated experiments, but there are well-defined distributions around the mean values. Both on the permutations and on the symbolic sequences, the refining process makes $C_{LZ}$ and $C_L$ separable: when order $r$ or $d$ is 1 or 2, $C_{LZ}$ and $C_L$ are almost equal; as the order increases the two measures separate in different directions until the critical order has been reached. At that time, the measures achieve the bounds given in Proposition 2 and its counterpart for permutations (see

Proposition 5 in Appendix B).

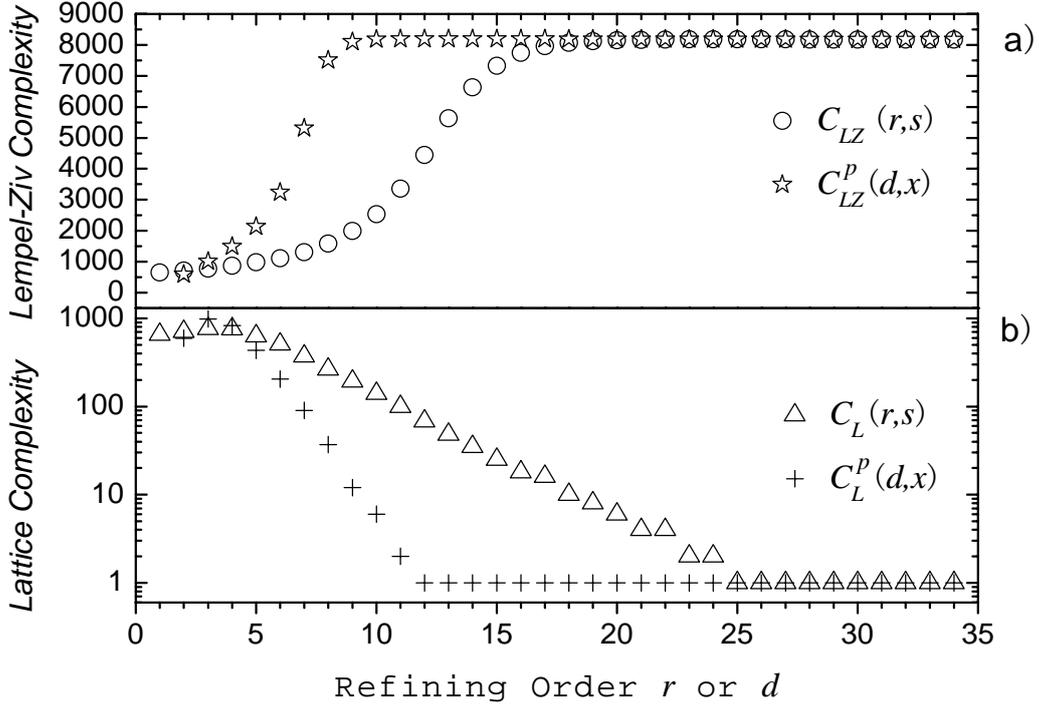

FIG.1. Complexity of a pseudo-random time series $x$ and a corresponding symbolic sequence $s$: (a) $C_{LZ}$ and $C_{LZ}^p$ versus the orders and (b) $C_L$ and $C_L^p$ versus the orders.

Similar results were obtained when the pseudorandom time series was created by other congruential generators including lagged Fibonacci generators [20].

Assuming the pseudo-random time series $x$ is of complete randomness, we may divide the whole range of the order, either $d$ or $r$, into a low region (about $d, r \leq 5$), a medium region ($5 < d < d^*$, $5 < r < r^*$) and a high region ($d \geq d^*$, $r \geq r^*$). The vertex of the maximum value of $C_L$ is in the low region and the low-order $C_L$ meets only criterion I. The transition from criterion I to II occurs within the medium region. Then $C_L$ behaves absolutely as a criterion-II measure in the high region. Nevertheless, what the values of the orders take as the boundaries of these regions depends on what a sequence is defined as the completely random matter, the "random zero."

Unfortunately, the completely random sequence has still not been clearly defined. The De Bruijn sequence may provide a convenient reference point of the "random zero." To get a criterion-II $C_L$, we must take the order $r > k$ for any symbolic sequence of length $n = \alpha^k + k - 1$. Since permutations are unsuitable for comparing with De Bruijn sequence, another kind of *emblematic sequence* should act as the approximation of the completely random case.

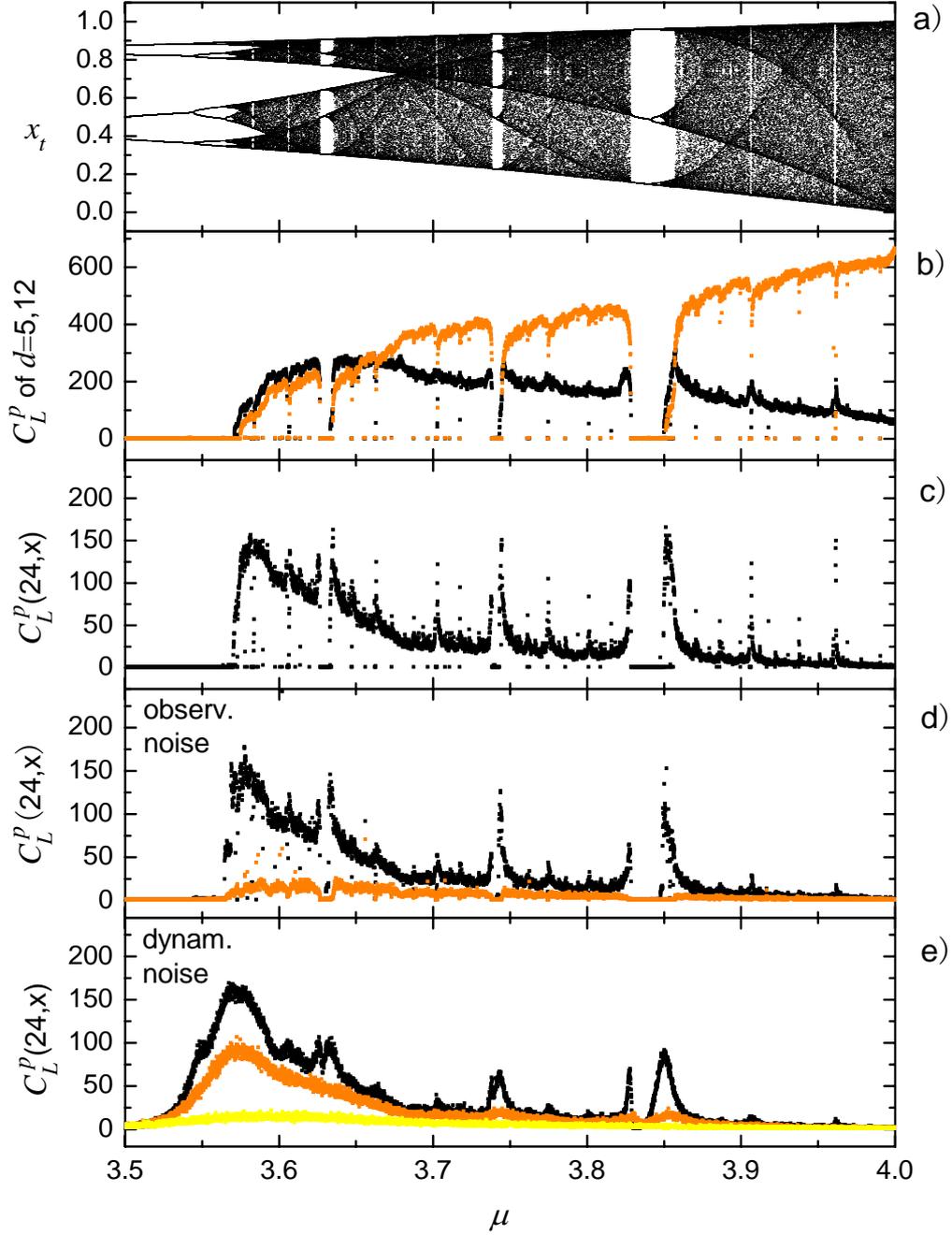

FIG.2. (Color online) Permutation lattice complexity $C_L^p$ for the logistic map as a function of $\mu$ (for each step of $\Delta\mu = 10^{-4}$, $n = 8204$, with the previous 25 000 points deleted as transient). (a) Bifurcation diagram. (b) $C_L^p(5, x)$ and $C_L^p(12, x)$ (the former increased globally across the chaotic region and intersected with the latter at about $\mu = 3.66$). (c) $C_L^p(24, x)$. (d) $C_L^p(24, x)$ with Gaussian observational noise at standard deviations $s = 0.00025$ (upper) and $s = 0.004$ (lower). (e) $C_L^p(24, x)$ with additive Gaussian dynamical noise at the standard deviations $s = 0.00025$, $s = 0.001$, and $s = 0.004$ (from upper to lower).

Except for the emblematic sequences, there are other more delicate options of "random zero": first, a pseudo-random time series as shown in Fig.1; second, the time series that have the most random look among the sample set (especially when you are familiar with the generating dynamics) as shown in the following instance.

To illustrate more properties of our measures, we took the logistic map $x_{t+1} = \mu x_t (1-x_t)$ with $3.5 \leq \mu \leq 4.0$ as another example. The Feigenbaum diagram is given in Fig.2(a) for comparison. Our measure was calculated with the order from 2 to 40. Since the time series for $\mu = 4$ is evidently the most chaotic one, we took it as the "random zero." The critical orders of different lengths $n$ are shown in Table 1. Fixing the length as $n = 8204$, the results show that the low-order ($d \leq 5$) $C_L$ matches criterion I and behaves like $C_{LZ}$ [15] and PE [4]. With medium orders ($5 < d < 24$), $C_L$ for $\mu = 4$ is no longer the maximum. With order 12, the critical order of the foregoing pseudo-random time series (Fig.1), $C_L$ seems more like a criterion-II measure [Fig.2(b)]. When the critical order 24 (see Table 1) of the "random zero" has been reached, $C_L$ becomes thoroughly a measure of criterion II [Fig.2(c)] and the case $\mu = 4$ is qualified as the simplest, as well as the periodic case now called "regular zero." The maximum value of $C_L$ is 166 at $\mu = 3.8512$.

High-order $C_L$ may attain the maximum values proximately at the edges of the chaotic regions. Some other criterion-II measures, fluctuation complexity for example, calculated by Wackerbauer *et al.* [21], showed similar results. Nevertheless, they have not been shown to qualify completely chaotic orbits as the simplest cases. They also do not exhibit the convenience of quick processing as $C_L$ has. Specifically, the outcome of $C_L$ can be decreasing with increasing $\mu$ within the chaotic region except the edges so that the different chaotic regions can be identified easily [Fig.2(c)].

As in PE, in our measure any strictly monotonic function on the original data cannot make any change. Such a property is crucial to physiological applications since the data may be collected with different equipments. In fact, all the merits of PE about resisting noise [4] are still preserved. Moreover, noise sometimes can even be useful for distinguishing chaotic and periodic regions. Since the equal values are sensitive to noise, for periodic orbits the permutations of a sufficiently high order (larger than the period) may engage so many deformations that show the complete randomness rather than the regularity. Unlike PE [see Fig.2(e) and (f) in Ref. [4]], $C_L$ in periodic region can remain still much lower than that in chaotic regions [Figs.2(d) and 2(e)], because high-order $C_L$ can approach the "random zeros" as well as the "regular zero."

In this instance, adding small random perturbations is equivalent to adding observational noise. If perturbations had been used to break the equality between the periodic points, there would exist some distortion in periodic regions in Fig.2(c) or 2(d) of Ref. [4].

**TABLE 1** Critical orders (of *permutations* and symbolic sequence) for varying lengths $n = 2^k + k - 1$ of time series of logistic map ($\mu = 4$). The first 25 000 steps were dropped and every data point was in double precision (64 bits total).

| $k$ | 7 | 8 | 9 | 10 | 11 | 12 | 13 | 14 | 15 | 16 |
|---|---|---|---|---|---|---|---|---|---|---|
| $n$ | 134 | 263 | 520 | 1033 | 2058 | 4107 | 8204 | 16397 | 32782 | 65551 |
| $d^*$ | 13 | 14 | 17 | 17 | 19 | 21 | 24 | 24 | 28 | 30 |
| $r^*$ | 14 | 16 | 19 | 19 | 21 | 22 | 26 | 26 | 31 | 34 |

Another advantage of $C_L$ contributing to real-world applications is that it does not need large data. A too long length of the time series may even cause troubles. First, larger data need longer time to calculate. Second, a longer length leads to higher critical order of the "random zero" (Table 1) and then still leads to a longer time to calculate. Even though $C_L$ is easy to calculate, when $n > 10^6$ and $r > 50$ or $d > 50$, the computing time may be a problem especially for a real-time process. On the other hand, a too short length will inevitably reduce the accuracy of the complexity measurements. It can be found that for a length of $n = 134$ and its corresponding proper order $d = 13$ (Table 1), the behavior of $C_L$ is still like that for longer length [Fig.2(c)]. However, the maximum $C_L$ becomes only 12.

When we use a smaller $\Delta\mu$ and take a short range of $\mu$, other kinds of crude copies of the aforementioned results can be obtained. For instance, we took the range $3.5647 \le \mu \le 3.6787$ containing the critical point $\mu = 3.57$. With $\Delta\mu = 2.5 \times 10^{-5}$ and $d = 38$ (the critical order of the most chaotic case $\mu = 3.6787$ in this situation), the diagram of $C_L$ is still much similar as shown in Fig.2(c). The maximum $C_L$ is 107 at the point $\mu = 3.5748$. This manifests not only the self-similarity of the map, but also the relativity and the adaptability of our measure.

## V. CONCLUSION

In summary, we have presented an easily adaptable complexity measure that meets almost all existing criteria. Notwithstanding the conditional agreement with criterion-I measures as $C_{LZ}$ and PE, our measure allows both the completely chaotic case and the periodic case to be regarded as the simplest object and allows the maximum outcome to be obtained near the "edge of chaos." By changing the control parameters, one can let $C_L$ simulate many other measures. However, our procedure is not designed to replace all the measures in use or to help to define the "one and only" criterion of complexity measure. The modified permutation method enables the measure to process all finite time series including that which has a large number of equal values. Monotonic transformations of the time series still have no effect on the permutations, and a certain degree of robustness against noise now is found not only for the chaotic signal, but also for the periodic signal. These enhancements hold promise for more extensive applications of $C_L$.

Besides the practical advantages of our measure, the results on finite time series negate the conventional view of Kolmogorov complexity that this definition merely interprets the randomness. The concepts including *critical order*, *iterative sequence* and *emblematic sequence* may shed lights on our understanding of complexity and randomness. Further work on the transformation between the symbolic sequences and the permutations may provide fundamentals of a conceivable applied permutation dynamics.

# APPENDIX A: THE REFINED GNERATING PARTIONS AND SYMBOLIC DYNAMICS

For the logistic map $x_{t+1} = \mu x_t (1 - x_t)$, the generating partition is the critical point $x_c = 1/2$. An iterative orbit starting from a point $x_0$ gives a unique symbolic itinerary $s = s_0 s_1 s_2 \cdots$, where $s_i = 0(1)$ if $f^i(x_0) < x_c (> x_c)$. So all the information of the point $x_0$ can be encoded as an infinite binary sequence, and every finite word of this sequence represents an interval covering a point visited by the orbit [16]. When the map is surjective (i.e. $\mu = 4$ as shown in Fig.3), there is no forbidden symbolic sequence, meaning that any sequence corresponds to an orbit.

In Fig.3, the segment (c1, c2) corresponds to the word 001 so that a point between c1 and c2 produces a symbolic sequence with 001 as the first three symbols. Suppose there is a finite sequence 1010010011 of length 10; we transform it into an eight-symbol refined sequence (101,010,100,001,010,100,001,011) by taking the order 3. We see that the corresponding orbit must visit eight points within five distinct intervals. The fourth and sixth points, for instance, are within the same interval (c1, c2), but we cannot see from the sequence whether these two points are really the same until we take a higher order to get more information of every point.

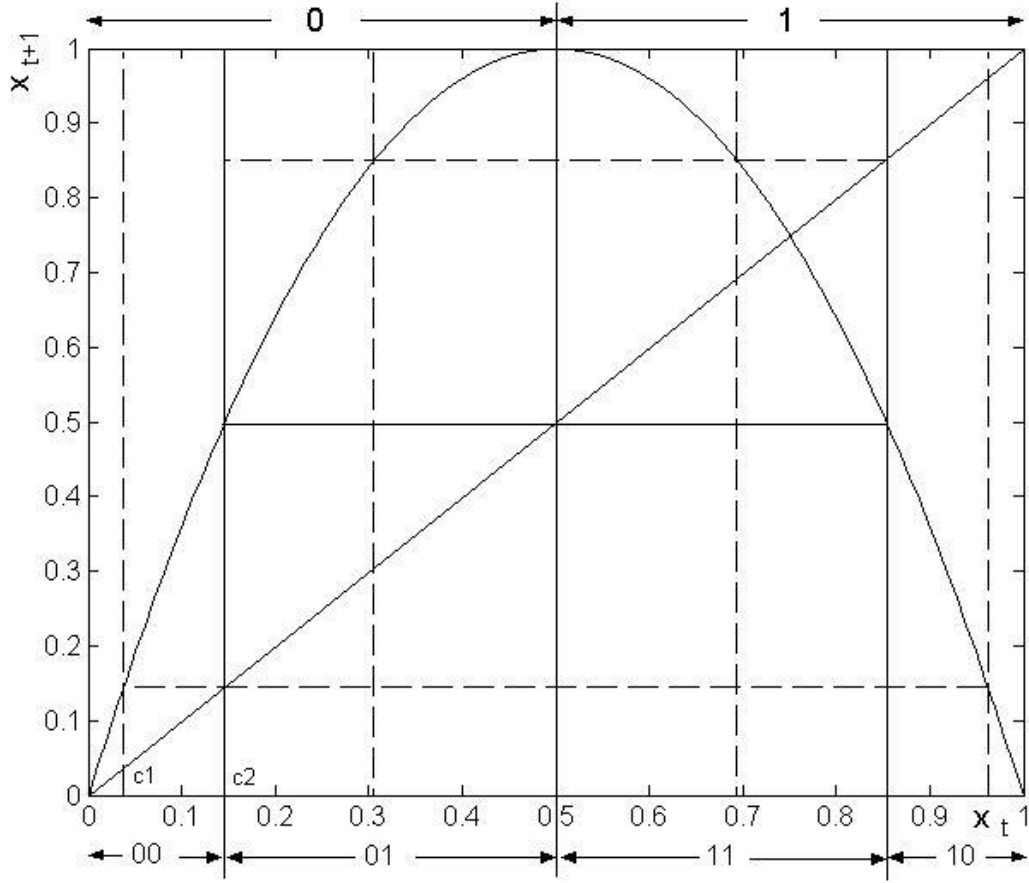

FIG.3. Logistic map and its generating partition with further refinement (the dashed lines represent the further critical points of the three-order refinement, dividing the whole phase place into eight intervals corresponding to 3-bit words).

Simple conventions can be used to order all symbolic sequences according to the order of their corresponding segments' position [17]. Let $0 < 1$ for 1-bit words. Any two words having the same prefix $\Sigma$ can be ordered by the conventions as follows:

$$\Sigma 1 \cdots > \Sigma 0 \cdots \text{ if } \Sigma \text{ has an even number of 1,}$$
$$\Sigma 1 \cdots < \Sigma 0 \cdots \text{ if } \Sigma \text{ has an odd number of 1.}$$

# APPENDIX B: THE PROPOSITIONS AND THEIR PROOFS

For an original time series $x$ of length $n$ and a given order $d$, let $C_{LZ}^{p}(d,n)$ and $C_{L}^{p}(d,n)$ denote *permutation* $C_{LZ}$ and *permutation* $C_{L}$ respectively. We may convert the forgoing three propositions in Sec. II into new forms as follows.

**Proposition 4.** When $x$ is an $m$-periodic time series ($m < n$), we have
$$C_{LZ}^{p}(m,n) = m,$$

$$C_L^p(m,n) = 1.$$

This proposition is obvious. It shows that any periodic time series can easily be regarded the simplest object by using *permutation* $C_L$.

**Proposition 5.** If $x$ satisfies the conditions that every point is distinct from the others and that every permutation of any given order $d$ has the uniform probability to appear, for $d \sim n$ we see that

$$\lim_{n \to \infty} \lim_{d \to \infty} \Pr[C_{LZ}^p(d,n) = n - d + 1] = 1,$$

$$\lim_{n \to \infty} \lim_{d \to \infty} \Pr[C_L^p(d,n) = 1] = 1.$$

**Proof.** Without losing generality, we may assume the probability distribution of $d$ permutations is $p(d)$, $0 < p(d) < 1$. Then a $d$ permutation has the probability $p(d) \times p(d) = p^2(d)$ to occur twice. We have $C_L^p(d,n) > 1$ if and only if there exist two or more of the same $d$ permutations with different next neighbors. There are $n - d + 1$ permutations of order $d$ in $x$; hence,

$$\Pr[C_L^p(d,n) > 1] \leq (n-d) p^2(d).$$

Because every $(d + 1)$ permutation has a uniform probability to appear and because there are $d + 1$ possible distinct $d + 1$ permutations extended from a $d$ permutation, we get

$$p(d+1) = p(d)/(d+1).$$

Since the least value of $d$ is 2 and $p(2) = 1/2$, by induction on $d$ we have $p(d) = 1/d!$ and then

$$\lim_{n \to \infty} \lim_{d \to \infty} \Pr[C_L^p(d,n) > 1] \leq \lim_{n \to \infty} \lim_{d \to \infty} \frac{(n-d)}{(d!)^2} = 0 \quad (d \sim n).$$

Because $C_L^p(d,n) \geq 1$ is always valid, we see that $\lim_{n \to \infty} \lim_{d \to \infty} \Pr[C_L^p(d,n) = 1] = 1$. This means the all $(n - d + 1)$ $d$ permutations are distinct with probability 1, so it is true that $\lim_{n \to \infty} \lim_{d \to \infty} \Pr[C_{LZ}^p(d,n) = n - d + 1] = 1$.

**Q.E.D**

For a time series of constant length $n$, the value of $C_L$ is affected by the order $d$. The propositions show that the periodic and random series will be both regarded as the simplest cases when we take a large enough order.

**Proposition 6.** If $x$ is a time series of length $n = l!+l-1$ with every $l$-permutation occurring just once, we have the permutation critical order

$$d^*(x) = l.$$

**Proof.** Directly from the definitions of $C_L$ and $x$, we see that $C_L^p(l,n) = 1$. If it is true that $C_L^p(d,n) = 1$ for any $d > l$ and $C_L^p(d,n) > 1$ for $d = l-1$, $l$ is the critical order.

i) When $d > l$, all $d$-permutations of $x$ are distinct, for otherwise there exist repeated $l$ permutations and we get a contradiction. Hence, for any $d > l$, $C_L^p(l,n) = 1$.

ii) When $d = l-1$, there are $(l-1)!$ different permutations of order $l-1$. At the same time $x$ exerts $l!+1$ permutations, implying that there exist some repeated $l-1$ permutations. Then the whole permutation sequence of $x$ cannot be regarded as a chaotic sequence. Nor is it a periodic sequence, for otherwise the last symbols of a pair of $l$-permutations extended from a pair of repeated $l-1$ permutations must be the same and this leads to a contradiction. Hence, $C_L^p(l-1,n) > 1$.

**Q.E.D**

Proposition 6 indicates that among all permutation sequences that have a uniform distribution of permutations, the permutation sequence extracted from the time series described above is the one with the smallest critical order.

______________________________________________________

However, anyone who gets adaptable measures in this manner has to take the risk of involving pure mathematical games unless physical or practical meanings behind these measures are clarified. Further details and some "new" measures are given in another paper, which has been submitted for publication [Ke and Tong, 2008].